\documentclass[preprint]{aastex}

\usepackage{bm}

\DeclareMathAlphabet{\tens}{OT1}{cmss}{bx}{n}

\begin{document}

\title{Improved search of PCA databases for spectro-polarimetric inversion}

\author{R.\ Casini,\altaffilmark{a}
A.\ Asensio Ramos,\altaffilmark{b}
B.\ W.\ Lites,\altaffilmark{a}
A.\ L\'opez Ariste\,\altaffilmark{c}}

\affil{%
\altaffilmark{a}%
High Altitude Observatory, National Center for Atmospheric
Research,\footnote{The National Center for Atmospheric Research is
sponsored by the National Science Foundation.}\break
P.\ O.\ Box 3000, 80307-3000, Boulder, CO, U.S.A.\break
\altaffilmark{b}%
Instituto de Astrof\'{\i}sica de Canarias, c/ V\'{\i}a
L\'actea s/n, E-38205 La Laguna, Tenerife, Spain\break
\altaffilmark{c}%
THEMIS, CNRS UPS 853, c/ V\'{\i}a L\'actea s/n, 
E-38200 La Laguna, Tenerife, Spain}

\begin{abstract}
We describe a simple technique for the acceleration of
spectro-polarimetric inversions based on principal component analysis
(PCA) of Stokes profiles. This technique involves the indexing of 
the database models based on the sign of the projections (PCA 
coefficients) of the first few relevant orders of principal components
of the four Stokes parameters. In this way, each model in the database 
can be attributed a 
distinctive binary number of $2^{4n}$ bits, where $n$ is the number 
of PCA orders used for the indexing. Each of these binary numbers 
(indexes) identifies a group of ``compatible'' models for the inversion 
of a given set of observed Stokes profiles sharing the same index. The 
complete set of the binary numbers so constructed evidently determines 
a partition of the database. The search of the database for the PCA 
inversion of spectro-polarimetric data can profit greatly from this 
indexing. In practical cases it becomes possible to approach the 
ideal acceleration factor of $2^{4n}$ as compared to the systematic 
search of a non-indexed database for a traditional PCA inversion. 
This indexing method relies on the existence of a physical meaning 
in the sign of the PCA coefficients of a model. For this reason,
the presence of model ambiguities and of spectro-polarimetric noise in the
observations limits in practice the number $n$ of relevant PCA orders
that can be used for the indexing.
\end{abstract}

\section{Introduction} \label{sec:intro}

One of the biggest challenges that the astronomical community faces with 
the next generation of astronomical instrumentation, both ground based 
and space borne, is without doubt the enormous amount of data that will
need to be stored, reduced, analyzed, and finally interpreted in terms 
of physical processes. This problem is particularly striking in the case 
of solar observations, where essentially every pixel of the detector 
contributes to the science data. As an example, the expected volume of
spectro-polarimetric data that will be produced by the Advanced
Technology Solar Telescope \citep{Ri08} is of the order of 16\,TB per
day.

The inversion of spectro-polarimetric data is by itself a notoriously 
challenging problem of solar physics. This is because the emergence of 
a polarized signal in the solar spectrum, starting from a fundamentally 
isotropic (and hence, unpolarized) radiation within the solar interior,
can only be explained through a complicated description of the 
interaction of light with a gas of ions in a temperature and pressure 
stratified atmosphere, and in the presence of magnetic fields that are
often entangled down to and below the smallest spatial scales observable
with present-day instrumentation. For this reason, the inversion of
spectro-polarimetric signals is intrinsically an ill-posed problem, and
the forward modeling of the polarized solar spectrum can be very time
consuming, depending on the type of spectral lines considered, and on
where these lines form on the Sun \citep{TB10,Ca12}.

The development of time-efficient inversion techniques for
spectro-polarimetric data is a thriving field of research in solar
physics. We will not discuss here the merits and issues of the various
approaches to spectro-polarimetric inversion, and we refer instead the 
reader to a review by \cite{AR12} for such a discussion. Here we 
want to focus on the problematics associated with inversion methods that
rely on pattern-recognition techniques, and more specifically,
\emph{principal component analysis} \citep[PCA;][]{Pe01,Jo02}, which
has successfully been applied to spectro-polarimetric observations of the
solar photosphere \citep{Re00}, and of solar prominences and filaments
\citep{LC03,Ca03,Ca05,Ku09}. 

In the next section we briefly summarize the ideas behind PCA-based
inversion of spectro-polarimetric data (see also, 
\citealt{Re00,SL02,LC02}). In Section~\ref{sec:strategy}, 
we present our idea for the indexing of the inversion database, 
its justification, and its limits of applicability.
Finally, in Section~\ref{sec:results}, we present some test results 
of PCA inversions with indexed databases, applied to on-disk 
observations of \ion{He}{1} 1083\,nm that were performed by one of 
us (B.L.).
We provide an outlook for further development in our concluding remarks.

\section{Principal Component Analysis} \label{sec:theory}

We briefly summarize below the basic 
concepts of PCA Stokes inversion, for the sake of completeness, and 
also for introducing notation essential to this study.

We consider the case of a spectrally and spatially resolved observation 
of a solar region with a spectro-polarimeter. This observation is fully 
characterized by the set of Stokes vectors 
$\bm{S}_j(\lambda_i)\equiv(I_j(\lambda_i),Q_j(\lambda_i),U_j(\lambda_i),
V_j(\lambda_i))$, where $i=1,\ldots,N$ indicates the wavelength points
sampled by the spectro-polarimeter, and $j=1,\ldots,M$ indicates the
spatially resolved elements of the observed region. In the Stokes
notation, $I$ refers to the light intensity, $Q$ and $U$ are the two 
independent states of linear polarization on the plane normal to the
direction of light propagation, and $V$ is the Stokes parameter for circular 
polarization around that direction. Let us indicate with $S$ any of 
the four Stokes parameters, $I$, $Q$, $U$, and $V$. Then the 
spectro-polarimetric observation of the solar region \emph{for that 
parameter} naturally define the $N\times M$ matrix
\begin{equation}
\tens{S}_{ij}=S_j(\lambda_i)\;,
\qquad i=1,\ldots,N\;;\quad j=1,\ldots,M\;.
\end{equation}

The $M$ observed points in the solar region form a set of statistically
independent realizations of the Stokes profile $S(\lambda)$. We can
then calculate the averages
\begin{equation}
\bar{S}(\lambda_i)=\frac{1}{M}\sum_{j=1}^M \tens{S}_{ij}\;,
\qquad i=1,\ldots,N\;,
\end{equation}
for each of the wavelength points, and the 
$N\times N$ \emph{covariance} matrix 
\begin{equation} \label{eq:covariance}
\tens{C}_{ij}=\sum_{l=1}^M
	[\tens{S}_{il}-\bar{S}(\lambda_i)]
	[\tens{S}_{jl}-\bar{S}(\lambda_j)]\;,
\qquad i,j=1,\ldots,N\;.
\end{equation}
This is a real and symmetric matrix, which therefore 
can always be diagonalized by an orthogonal transformation 
\citep[e.g.,][]{BM53}.
The solution of the corresponding eigenvalue problem,
\begin{equation} \label{eq:eigenproblem}
\tens{C}\bm{f}^{(k)} = e^{(k)}\bm{f}^{(k)}\;,
\qquad k=1,\ldots,N\;,
\end{equation}
is known to provide an optimal set of orthogonal 
\emph{eigenprofiles} -- represented
by the $N$-dimensional eigenvectors $\bm{f}^{(k)}$ -- for the decomposition of 
the residual signals $S_j(\lambda)-\bar{S}(\lambda)$ \citep{Jo02}.
These eigenprofiles are also known as the \emph{principal components} of
the observed set of profiles. 
Another property of the covariance matrix is to be positive
semidefinite \citep[e.g.,][]{Jo02}, hence $e^{(k)}\ge 0$, for all $k$.
In particular, solving the eigenvalue
problem (\ref{eq:eigenproblem}) by singular value decomposition (SVD;
e.g., \citealt{Pr07}) provides us with an ordered set of eigenprofiles
according to the decreasing non-negative amplitude
%\footnote{%
%Equation~(\ref{eq:covariance}) shows that the covariance matrix has
%the form $\tens{C}=\tens{M}\tens{M}^\mathrm{T}$, for an appropriate 
%matrix $\tens{M}$, so $\tens{C}_{ii}=\|\tens{M}_i\|^2\ge 0$, where
%$\tens{M}_i$ is the $i$-th row-vector of $\tens{M}$. If then
%$\tens{Q}=\tens{X}\tens{C}\tens{X}^\mathrm{T}
%	=(\tens{X}\tens{M})(\tens{X}\tens{M})^\mathrm{T}$ 
%is a diagonal form of $\tens{C}$, it is also
%$e^{(i)}=\tens{Q}_{ii}=\|(\tens{XM})_i\|^2\ge 0$.}
of the corresponding singular values.
This ordering reflects the importance of the contribution of the various
eigenprofiles to the covariance of the observations.

The eigenprofiles $\bm{f}^{(k)}$ form a basis for the space of the 
residual signals $S_j(\lambda)-\bar{S}(\lambda)$. In
particular, this implies that the $j$-th profile in the set of $M$ 
observations can be reconstructed \emph{exactly} from its set of \emph{PCA 
components},
\begin{equation} \label{eq:PCAcoeff}
c_j^{(k)}\equiv\sum_{i=1}^N \bm{f}_i^{(k)}[\tens{S}_{ij}-\bar{S}(\lambda_i)]\;,
\qquad k=1,\ldots,N\;,
\end{equation}
so that
\begin{equation} \label{eq:expansion}
S_j(\lambda)-\bar{S}(\lambda) \doteqdot
	\sum_{k=1}^N c_j^{(k)}\bm{f}^{(k)}\;.
\end{equation}
When the eigenprofiles $\bm{f}^{(k)}$ are ordered according to their
corresponding singular values, then any truncation of the summation in 
Equation~(\ref{eq:expansion}) provides an approximation of the residual 
$S_j(\lambda)-\bar{S}(\lambda)$. It is found in practice that a 
small number $n$ of eigenprofiles ($n\sim 10$) is often sufficient
to reconstruct the residual signals within the typical polarimetric noise 
of the observations.

\begin{figure}[p!]
\centering
\includegraphics[width=\hsize]{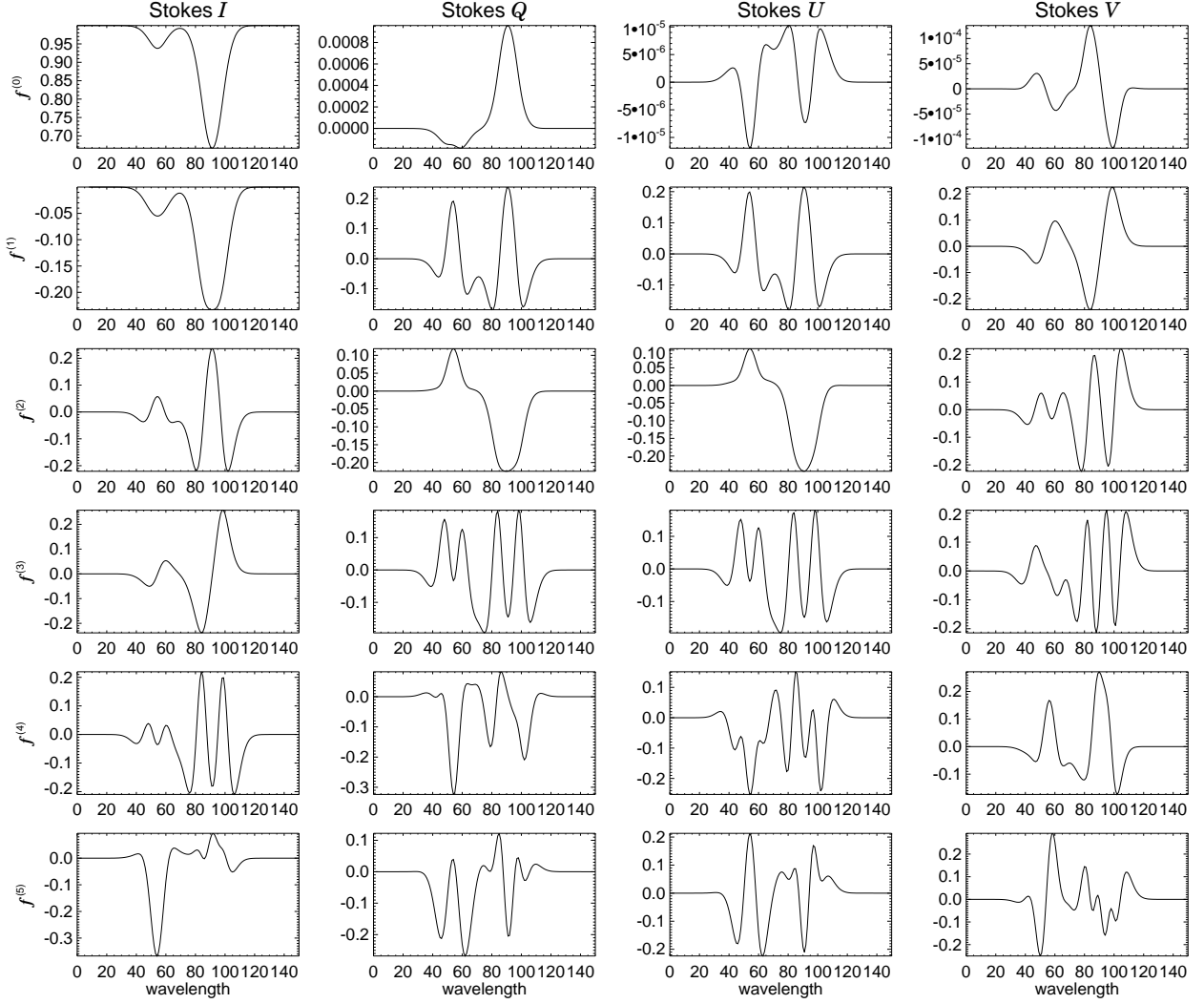}
\caption{\label{fig:eigenprofiles}
The average profiles (top row) and the first five PCA eigenprofiles 
(next five rows) for each of
the four Stokes parameters, $I$, $Q$, $U$, and $V$ (columns),
extracted from a database of synthetic Stokes profiles of the 
\ion{He}{1} chromospheric lines at 1083\,nm. The rows correspond to 
increasing orders of the eigenprofiles, with 
$\bm{f}^{(0)}=\bar{S}(\lambda)$.
The forward model used in the derivation of this eigenbasis corresponds 
to on-disk observations of the \ion{He}{1} lines, for inclinations of the 
line of sight to the local normal to the surface between $30^\circ$ and 
$40^\circ$. 
The database comprises all possible orientations of the magnetic field 
vector (i.e., $0\le\vartheta_B\le\pi$ and $-\pi\le\varphi_B\le\pi$), 
and magnetic strengths between 0.2 and 2000\,G logarithmically 
sampled. The calculated
profiles emerge from a homogenous slab of plasma with optical depth 
at line center between 0.1 and 1.5, and temperature between 
5000 and 25,000\,K. The position height of the slab (which
affects the radiation anisotropy) varies between 0 and 0.06\,$R_\odot$.
Bulk velocities are accounted for by randomly displacing the rest 
frequency of each model within a pixel unit of Doppler shift. Because of
the homogenous slab assumption, velocity gradients are not taken into
account.}
\end{figure}

\begin{figure}[ht!]
\centering
\includegraphics[width=.69\hsize]{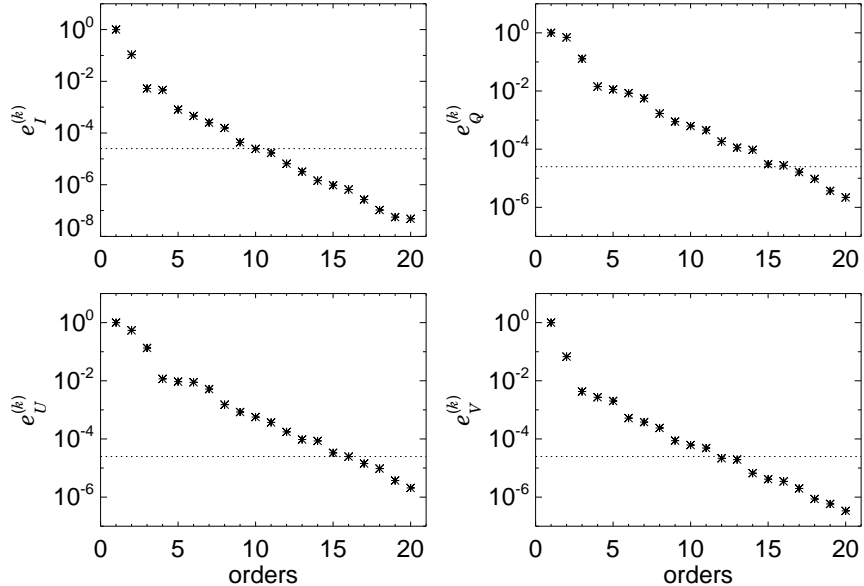}
\caption{\label{fig:eigenvalues}
Amplitudes of the first 20 singular values corresponding to the basis 
of eigenprofiles partially shown in Figure~\ref{fig:eigenprofiles}, 
for the four Stokes parameters, $I$, $Q$, $U$, and $V$. 
For each plot, the displayed values were normalized to the 
value for $k=1$. These amplitudes estimate the covariance of the model 
Stokes profiles along the ``principal directions'' represented by the 
eigenprofiles. The dotted lines show a threshold of
$2.5\times10^{-5}$, which corresponds approximately to an error of
$10^{-3}$ in the reconstruction of a synthetic profile (see text).}
\end{figure}

For the inversion of spectro-polarimetric data we must rely on a model
of the magnetized atmosphere that describes the line formation region.
Thus it is often preferrable to determine the eigenprofiles $\bm{f}^{(k)}$ 
from a database of synthetic line profiles based on the model
assumptions, rather than from the actual observations. Since they 
come from a model, these profiles are generally noiseless.
Figure~\ref{fig:eigenprofiles} show the first five eigenprofiles plus
the average profile, for each of Stokes~$I$, $Q$, $U$, and $V$, for the
multiplet of \ion{He}{1} at 1083\,nm observed on the disk. The figure 
caption lists the
ranges of the physical parameters for the model adopted for the
synthesis. This model is the same as the one used for the inversion tests
presented in Section~\ref{sec:results}. The set of synthetic 
profiles from which this eigenbasis is extracted consists of 50,000
models spanning the parameter space. In order to improve the sampling
``efficiency'' of such a limited number of models, we adopted the
\emph{Latin Hypercube Sampling} variant of the Monte Carlo method
\citep{MK79} for the construction of this set.

Figure~\ref{fig:eigenvalues} demonstrates instead the drop-off of the singular
values of the covariance matrix (\ref{eq:covariance}) for the first 
20 eigenprofiles. The relative magnitude of these singular values
estimates the model's covariance along any of the ``principal 
directions'' in the $N$-dimensional space spanned by the eigenprofiles.
From Figure~\ref{fig:eigenprofiles} we see that the peak amplitude of
the eigenprofiles is typically around 0.2. Thus, for a polarimetric 
noise of $10^{-3}$, we expect to be sensitive to profile covariances of 
the order of $(10^{-3}/0.2)^2=2.5\times10^{-5}$, which is
indicated in Figure~\ref{fig:eigenvalues} by the dotted line. From
that figure, taking also into account the local drop-off of the covariances 
around the specified threshold, we can conclude that, for the purpose of 
Stokes profile reconstruction and inversion, we should retain approximately 
11 orders for Stokes~$I$, 14 for $Q$ and $U$, and 13 for $V$. These
numbers must be compared with the dimension $N$ of the complete set of 
eigenprofiles in the database, which corresponds to the number of
wavelength points (in this case, $N=151$) used for the synthesis of 
the Stokes profiles. As a result, the description of the Stokes profiles
for our model in terms of their principal components allows a data 
compression of the spectro-polarimetric information by approximately
a factor 10.

It is important to observe that the above argument about the number of 
orders that must be retained for spectro-polarimetric inversions relies 
on two fundamental assumptions. The first assumption is that the error 
bars of the profiles are dominated by photon noise, and the second one 
that the photon counts is large enough that the associated poissonian 
noise can be treated as a random variate, so that its variance can
simply be added to that of the model. On the other hand, the systematic 
errors due to deviations of the observations from the line formation 
model, especially in the case of complicated atmospheric structures, 
is very likely to dominate the inversion errors. Thus, in practical 
cases, we should not expect that retaining such a high number of orders 
necessarily improves the goodness of the profile fits from the inversion.
We will come back to this argument in the next section.

\begin{figure}[t!]
\centering
\includegraphics[width=\hsize]{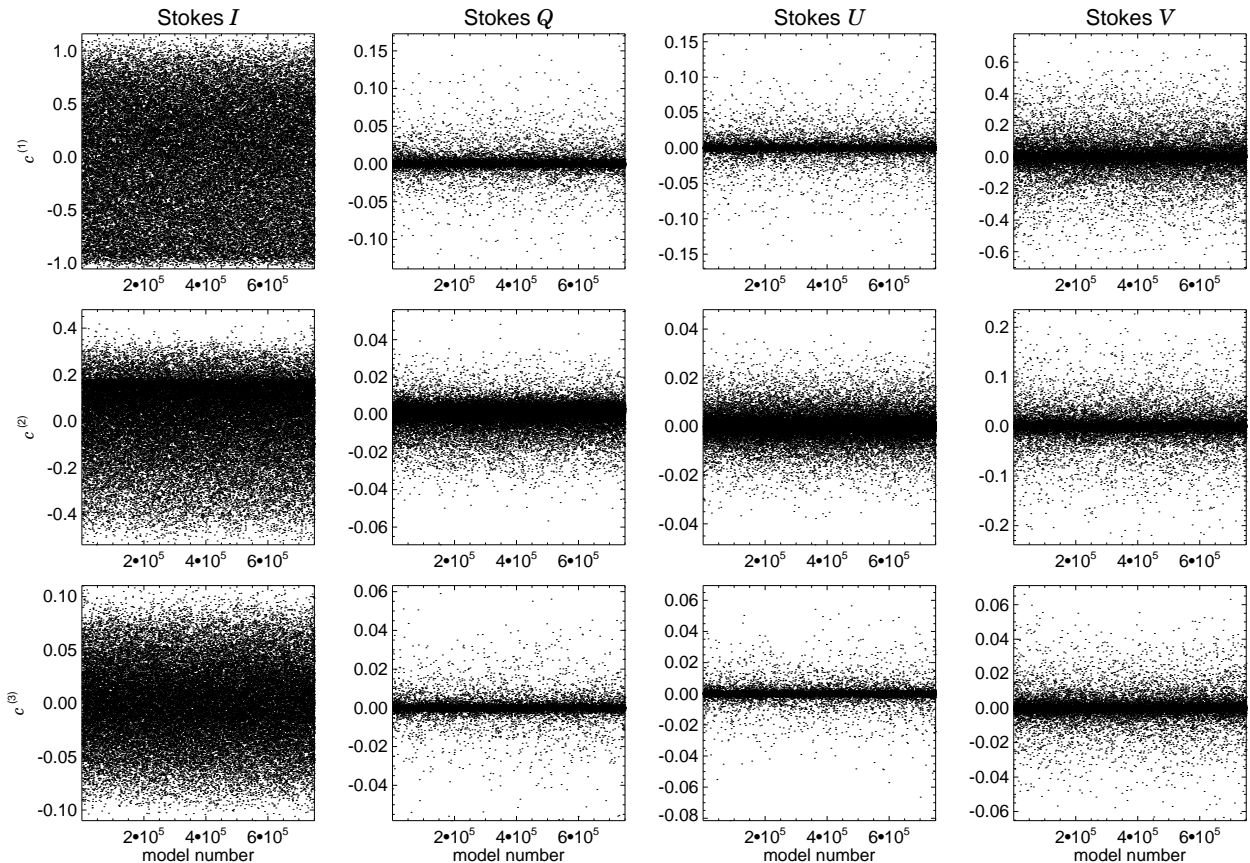}
\caption{\label{fig:coeffplot}
Scatter plots of the PCA coefficients for the atmospheric model 
considered in Figure~\ref{fig:eigenprofiles}. Only the first three 
orders of eigenprofiles (rows) for each of the four Stokes parameters, 
$I$, $Q$, $U$, and $V$ (columns), are shown. These plots span 
the entire inversion database, with a total of 0.75 million models. 
(For plotting convenience, here we only show one every fifteen points 
in the database.) We note the clear tendency for the PCA coefficients 
to be distributed rather evenly around a zero average.}
\end{figure}

\begin{figure}[p!]
\centering
\includegraphics[width=.69\hsize]{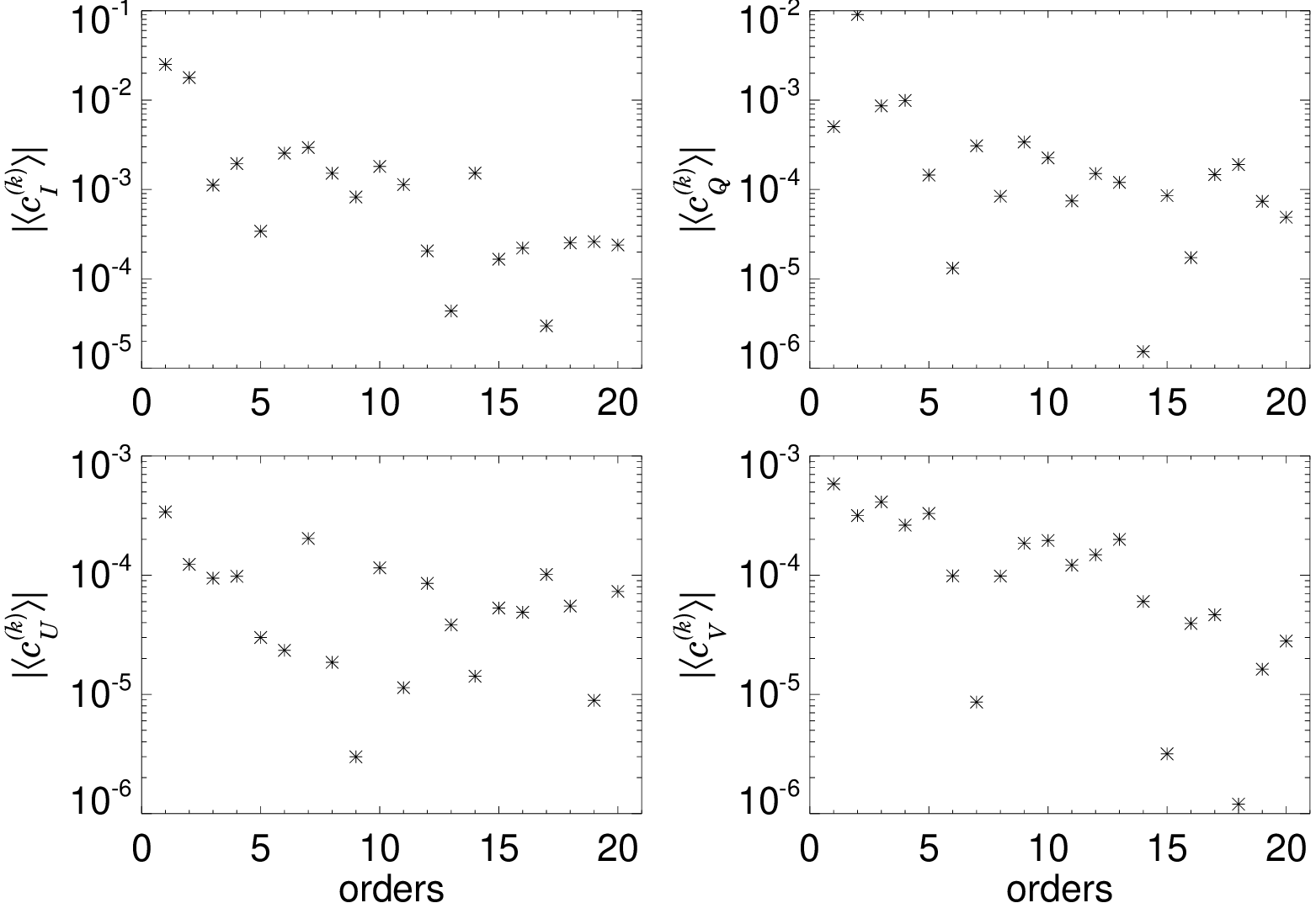}\vspace{8pt}
\includegraphics[width=.69\hsize]{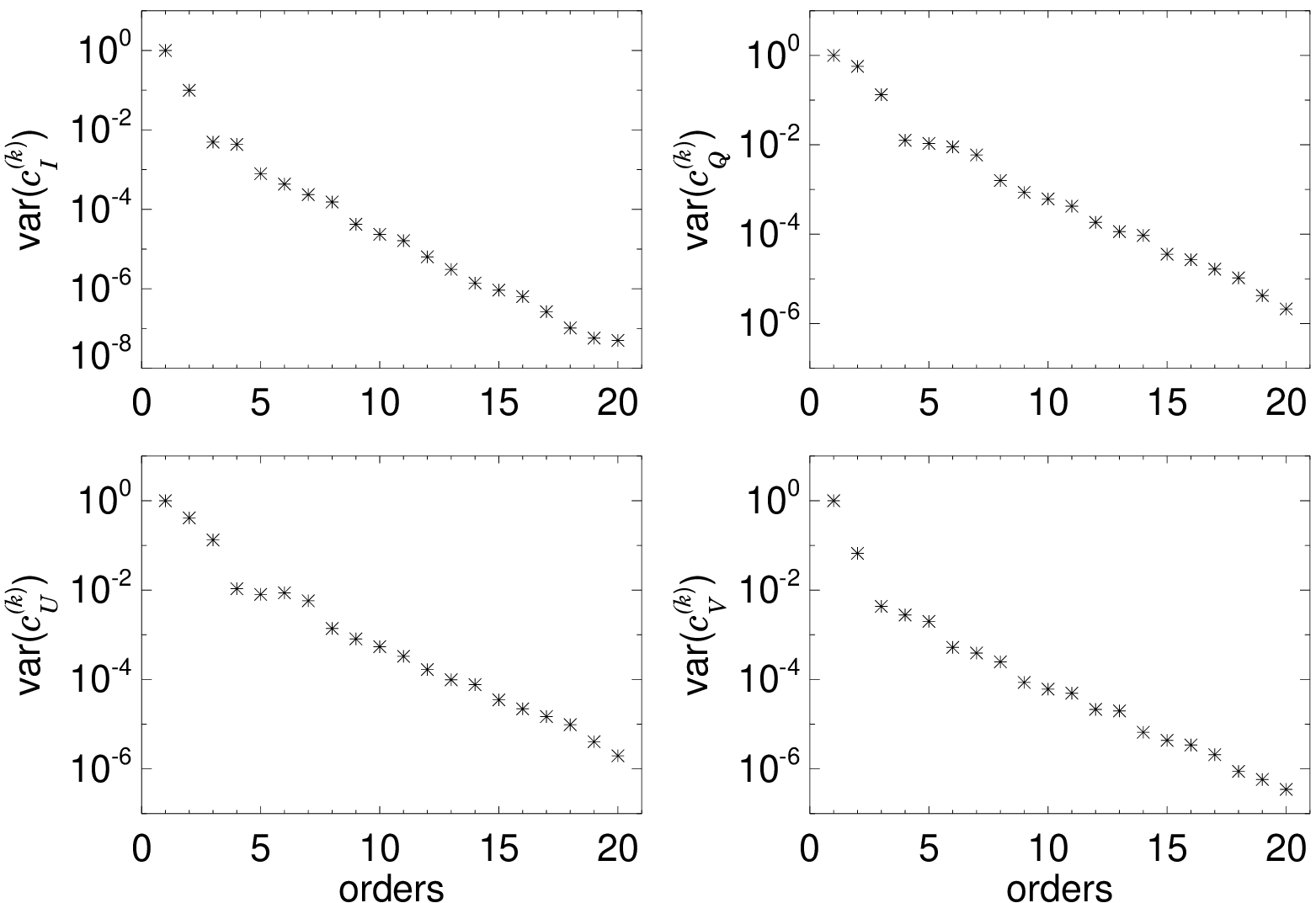}
\caption{\label{fig:meanvar}
Amplitudes of the absolute mean (top) and variance (bottom) of the first 
20 PCA coefficients for the four Stokes parameters, $I$, $Q$, $U$, 
and $V$, calculated over the same 50,000 models of the inversion 
database as for the case of Figure~\ref{fig:coeffplot}.
For each Stokes parameter and order $k$, the mean value of the 
PCA coefficient was normalized to its maximum absolute value within
the database. The first three points in each plot 
correspond then to the (normalized) mean values of the distributions 
shown in Figure~\ref{fig:coeffplot}. These plots confirm the fact 
that the PCA coefficients tend to be approximately distributed around 
zero. The variance values were instead normalized to the value for $k=1$.
As expected, the variance plots are practically identical to the plots
of the singular values of the covariance matrix, given in
Figure~\ref{fig:eigenvalues}.}
\end{figure}

\section{Indexing of PCA inversion databases} \label{sec:strategy}

We created an inversion database of 0.75 million models spanning the same
parameter space as the eigenbasis of Figure~\ref{fig:eigenprofiles}. The
profile information in this database is encoded in the expansion
coefficients given by Equation~(\ref{eq:PCAcoeff}). The inversion
database is constructed by a strategy of ``filtered'' 
Monte Carlo sampling, where each new randomly selected point in the 
parameter space is tested for proximity to previously included models 
in the database.
The testing parameter is the \emph{PCA distance} between two models, 
$i$ and $j$, which is defined as follows, for each of the four Stokes 
parameters:
\begin{equation} \label{eq:PCAdist}
d_{ij}=\left(\sum_{k=1}^m
\left[c_i^{(k)}-c_j^{(k)}\right]^2\right)^{1/2}\;,
\end{equation}
where $m$ is the maximum number of orders retained for the
reconstruction of the Stokes profiles.
The filtering criterion is to reject models for which the cumulative 
PCA distance for the four Stokes parameters\footnote{This cumulative
distance can be defined is several different ways. Its specific form
is not essential for the following discussion.} is less than a 
predefined value, $\delta$. The value of $\delta$ is reduced (increased) 
during the construction of the database when the rejection rate of new 
models becomes too large (small). This strategy allows to build up 
large-size databases where the shapes and amplitudes of the Stokes 
profiles are homogeneously distributed.

Because of the Monte Carlo construction of the inversion database, the
distribution of the models in the database is completely random. Thus
the PCA inversion ordinarily requires a full search through the database 
in order to identify the closest model (in terms of PCA distance) to the
observations, and this search must be repeated for each of the observed
set of Stokes profiles. For complicated atmospheric models, which may
depend on many parameters, the number of models in the database that is 
needed for an accurate PCA inversion can be in the order of millions. The
systematic search of such large databases represents one of the most
critical downsides of inversion methods based on pattern-recognition 
techniques. It is thus important to devise ordering strategies for the
inversion databases that can significantly reduce the search time.

The strategy that we propose in this work relies on the characteristic
distribution of the values of the PCA coefficients $c_i^{(k)}$ within 
the inversion database, for each of the four Stokes parameters. This 
distribution, for the 
atmospheric model and inversion database considered in this work, is 
presented in Figure~\ref{fig:coeffplot}, for $k=1,2,3$, and for the 
first 50,000 models of the database. We note immediately 
the tendency for the PCA coefficients to be distributed rather 
evenly around a zero average (see also Figure~\ref{fig:meanvar}). 
This is largely a consequence of the fact that the average Stokes
profiles are subtracted out for the definition of the covariance
matrix, Equation~(\ref{eq:covariance}).
Since the low-order eigenprofiles contribute more importantly 
to the reconstructed profiles (see the variance plots of 
Figure~\ref{fig:meanvar}, as well as Figure~\ref{fig:eigenvalues}), 
we can expect that the \emph{sign} of the PCA coefficients for these 
orders will also be important in determining the shape and amplitude of 
the Stokes profiles that must be matched to the observations. In 
other words, we 
can make the assumption that, \emph{in comparing models with
observations, the signs of the corresponding PCA coefficients for 
the lowest orders must match}.

Based on this assumption, we can partition the inversion database into
disjoint classes, each of them being characterized by a unique string of
signs characterizing the PCA coefficients of the models in that class.
In particular, we may convene to attribute the value 0 to ``$-$'' and 
the value 1 to ``$+$'', in which case each of these classes are 
identified by a unique binary number. Since there are four Stokes
parameters, each PCA order gets associated with a 4-bit number, and the
number of classes that each order brings to the partitioning corresponds 
to the number of integers that can be built with a 4-bit number, i.e.,
$2^4=16$. The total number of classes of the database partition is thus
$2^{4n}$, where $n$ is the number of orders used for the indexing.
The resulting partition classes can then be ordered according to 
the increasing value of the binary index number associated with
each class. If one class corresponds to a string of signs that have no match 
within the database, then that particular class will be empty.
This indexing of the database models allows to access directly the
desired class for the inversion. In practice, one determines the
indexing number from the PCA coefficients of the observations, and
then looks for the best matching model in the database by restricting
the search to the class with the same index. 

Typically, the order of
indexing, $n$, will be a small number, say, between 1 and 3. There are 
essentially two reasons for this. First of all, a high
level of partitioning could make the number of models in each class 
too small, depending on the total number of models in the database,
thus affecting the statistical significance of the inversion. Secondly,
because the amplitude of the PCA coefficients decreases rather rapidly 
with the order number (see variance plots of Figure~\ref{fig:meanvar}), 
the signs of high-order coefficients \emph{for the observations} 
could largely be affected by noise as well as other systematic errors. 

We already hinted to this problem at the end of the previous 
section. However, through the variance plots of Figure~\ref{fig:meanvar} 
we can better assess the relative significance of the high-order 
eigenprofiles for Stokes inversion. In fact, we can imagine to produce 
similar variance plots for the PCA coefficients of the decomposition of 
the observed data into their principal components. If the line formation 
model were representative of the observations, we should then expect 
that the variance plots for the observations and the database would be 
similar. In contrast, the presence of systematic effects in the 
observations, not accounted for by the model, would be demonstrated by 
a deviation between the two sets of variance curves, respectively, 
for the observations and the model. Typically, the two sets of variance 
curves will overlap for the very first PCA orders, and then start 
diverging from some order $k_0$ on. It then becomes meaningless -- and, 
in fact, even detrimental -- to retain any PCA order larger than $k_0$ 
for the inversion.

We can therefore conclude that high orders should not be trusted for 
the purpose of preselecting inversion classes within the database. 
This conclusion is supported also by the fact that the relative number 
of empty classes within the database is empirically found to increase 
with the number 
of orders used for the indexing. So there is an increased risk that an 
observed set of Stokes profiles, perhaps because of unknown systematic 
effects in the formation of the spectral line affecting one of these 
high-order signs, be matched to an empty class of the database and 
not be inverted.

%\textbf{%
Of course, it is possible to devise alternative strategies for the 
indexing of the inversion database. 
For example, in a study of the PCA inversion of the photospheric 
lines of \ion{Fe}{1} at 630\,nm observed in active regions, \cite{Ey05} 
partitioned the inversion database into three subsets, corresponding to 
the three distinct magnetic regions of sunspot umbra, sunspot penumbra, 
and quiet Sun. The charateristic magnetic fields and thermodynamics of 
those three photospheric regions allowed the computation of synthetic 
profiles using correspondingly different sets of model parameters and 
eigenprofiles. In that work, the purpose of the authors was not directly 
to improve the inversion speed, but rather to attain a denser covering 
of the parameter space for each of the identified photospheric regions, 
in order to improve the quality of interpolate inversions.%}

%\textbf{%
In our work, we focused instead on the distribution of the Stokes 
profiles' shapes as characterized by the distribution of the PCA 
coefficients within the database, rather than on the physical 
characteristics of the emitting solar region (although these two 
viewpoints are evidently correlated). The proposed method of database 
partitioning is obviously 
the simplest that can be devised, as it relies only on the assumption 
that the PCA coefficients' mean gathers around zero.%} 
This assumption is well justified, for the particular line 
formation model of the \ion{He}{1} 1083\,nm that we considered in this 
work, but it may not always be the case for other line formation models, 
for example, in a database created for near-limb data, where on-disk
absorption profiles and off-limb emission profiles may be mixed within
the database. Indeed, already in the database we considered for this 
work, we can notice some visible deviation from zero mean, as well as 
some skewness, in the distributions of the first two PCA coefficients 
for Stokes $I$ (see Figure~\ref{fig:coeffplot}). Our method could 
then be immediately generalized by adopting instead the true mean of 
a PCA coefficient's distribution, or otherwise its median. In 
particular, using the median would guarantee a better balanced 
population of the partition classes, with two immediate advantages: 
(1) overall improvement of the statistical significance of the 
indexed inversions, and (2) approaching of the increase factor 
of the inversion speed to the theoretical maximum of $2^{4n}$.
A further improvement of the indexing strategy for PCA databases 
would be to look at higher-order moments of the coefficient 
distributions, such as the variance. This information could be used 
to define a new partition class of models, where the distance of a 
given PCA coefficient from the mean is less than some prescribed 
fraction of a standard deviation. With this type of partitioning, 
the number of classes becomes $3^{4n}$, and so one can attain a 
reduction of the inversion time by two orders of magnitude already 
for $n=1$.

\begin{figure}[p!]
\centering
\includegraphics[width=.8\hsize]{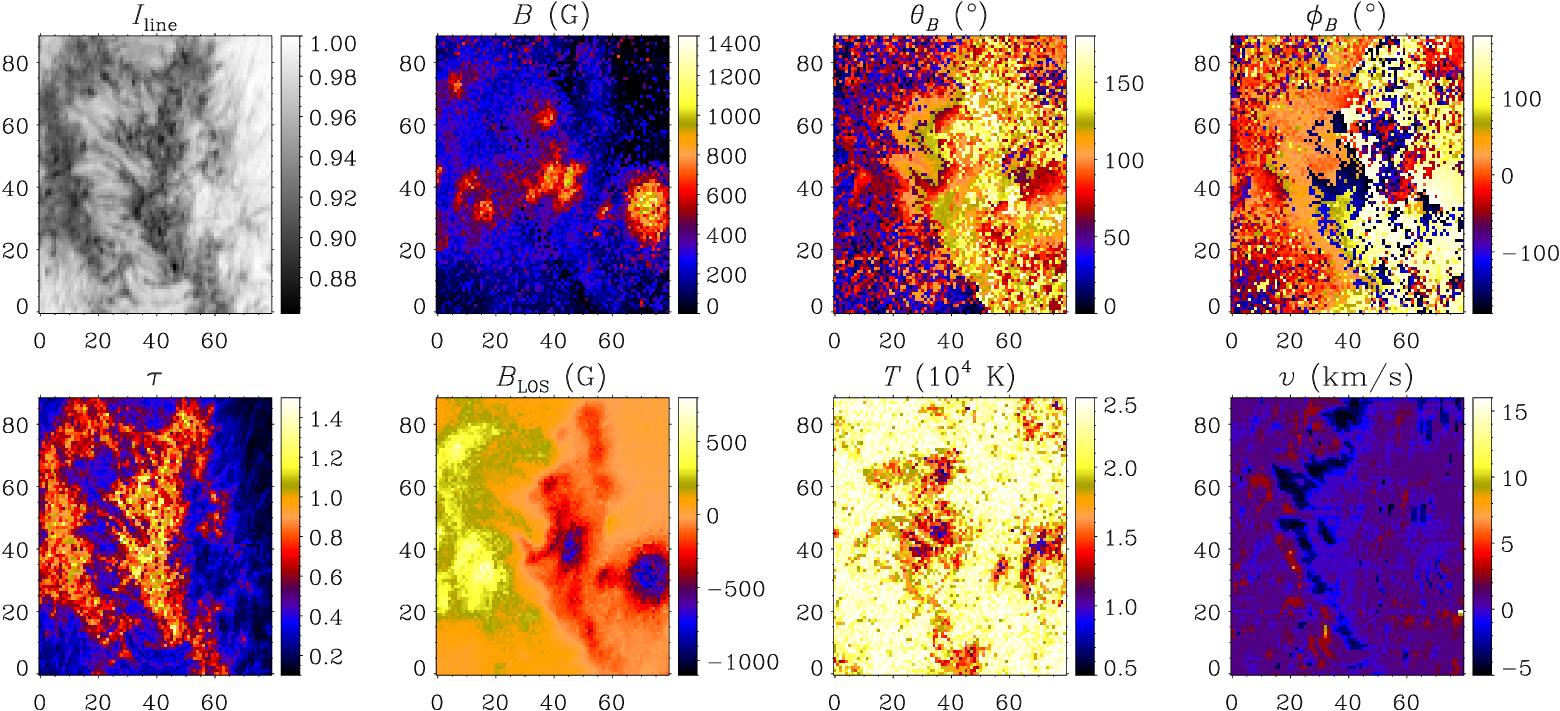}\vspace{8pt}
\includegraphics[width=.8\hsize]{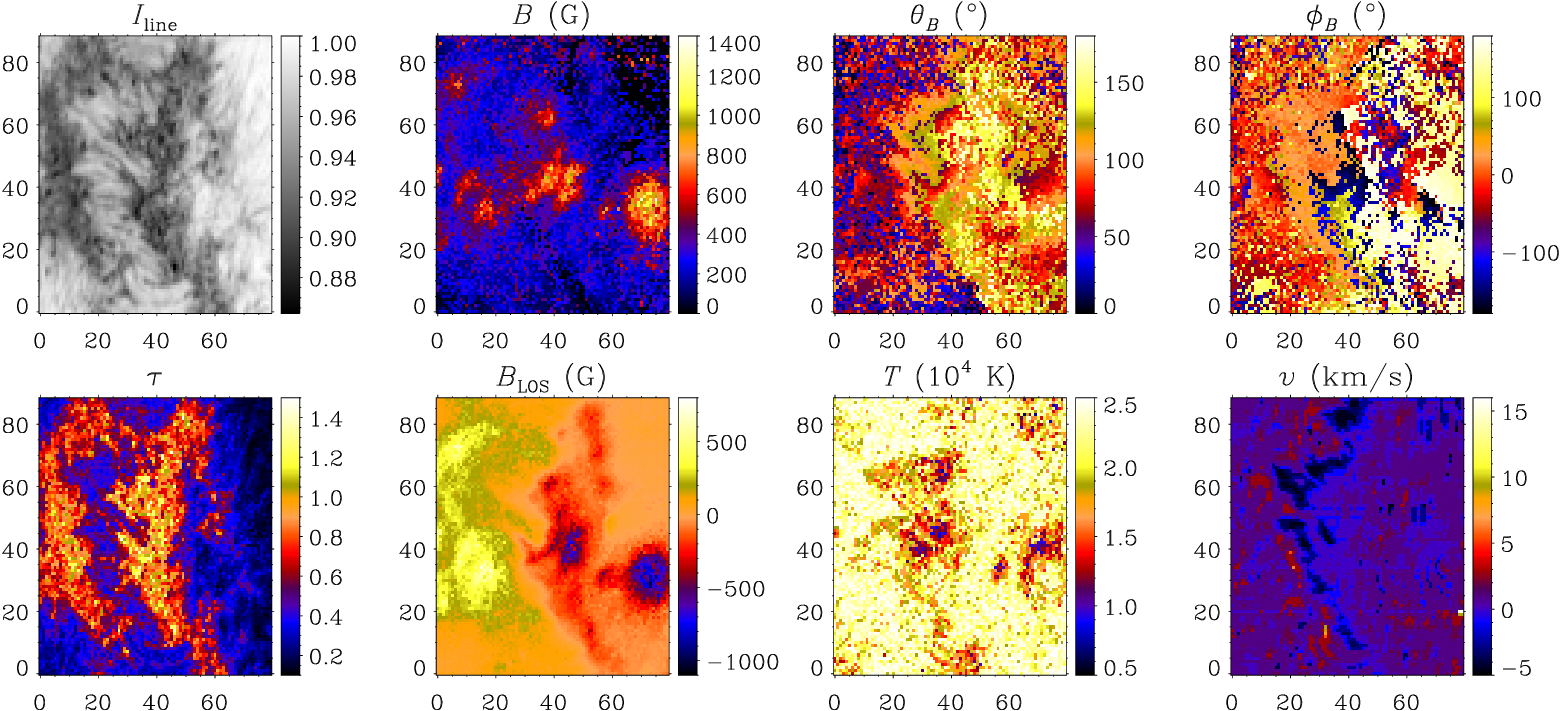}\vspace{8pt}
\includegraphics[width=.8\hsize]{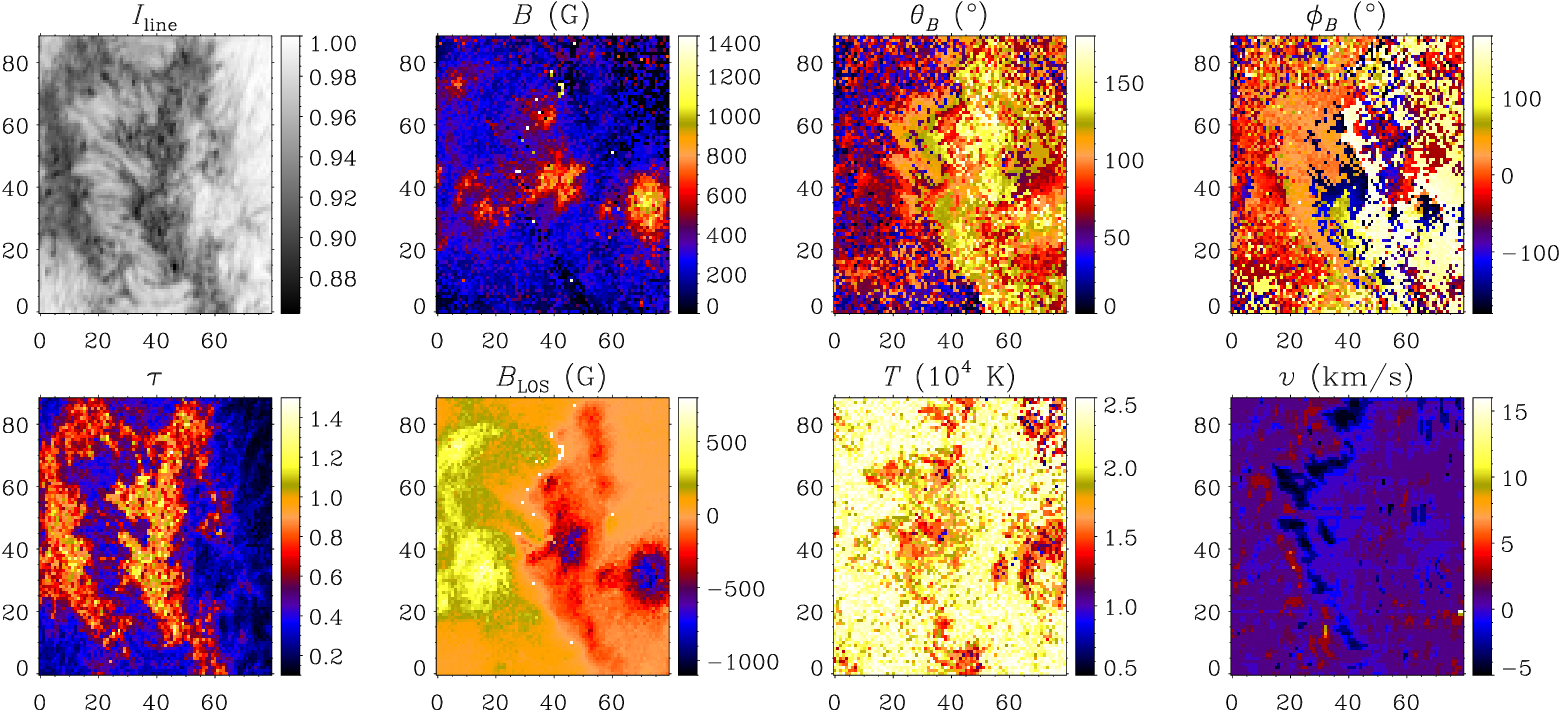}
\caption{\label{fig:results}
Magnetic maps for the dataset described in the text, corresponding to
inversions run with the original non-indexed database (top), and with 
indexed databases through the first order (16 partitions; 
middle) and second order (256 partitions; bottom) of eigenprofiles. 
See the text for a description of the inverted quantities.}
\end{figure}

\begin{table}[t!]
\centering
\begin{tabular}[c]{c|c|r}
\hline
indexing & reading time & inversion time \\
\hline\hline
none & 36\,s & 730\,s \hphantom{(0)} \\
1st-order (16 classes) & 38\,s & 50\,s (0) \\
2nd-order (256 classes) & 38\,s & 6\,s (0) \\
%no indexing & 30.3\,s & 1967.4\,s \\
%1st-order indexing & 32.4\,s & 158.6\,s \\
%2nd-order indexing & 32.8\,s & 29.4\,s (3) \\
\hline
\end{tabular}
\caption{\label{tab:results}
Inversion times for the magnetic maps of Figure~\ref{fig:results},
for three different orders of indexing of the PCA inversion database
using 0.5 million models out of the total 0.75 million in the 
database. Each inversion was run as a single thread on
a processor Intel Quad Core i7 2.2\,GHz. 
The increase factors of the inversion speed for the two cases of indexed
databases are approximately 15 and 120, respectively.
The number between 
parentheses next to the inversion time represents the number
of non-inverted points in the map for the indexed inversion.
The reading time is the overhead time necessary to read in 
the database and store its information in the computer memory for the 
inversion.}
\end{table}

\begin{figure}[t!]
\centering
\includegraphics[width=.8\hsize]{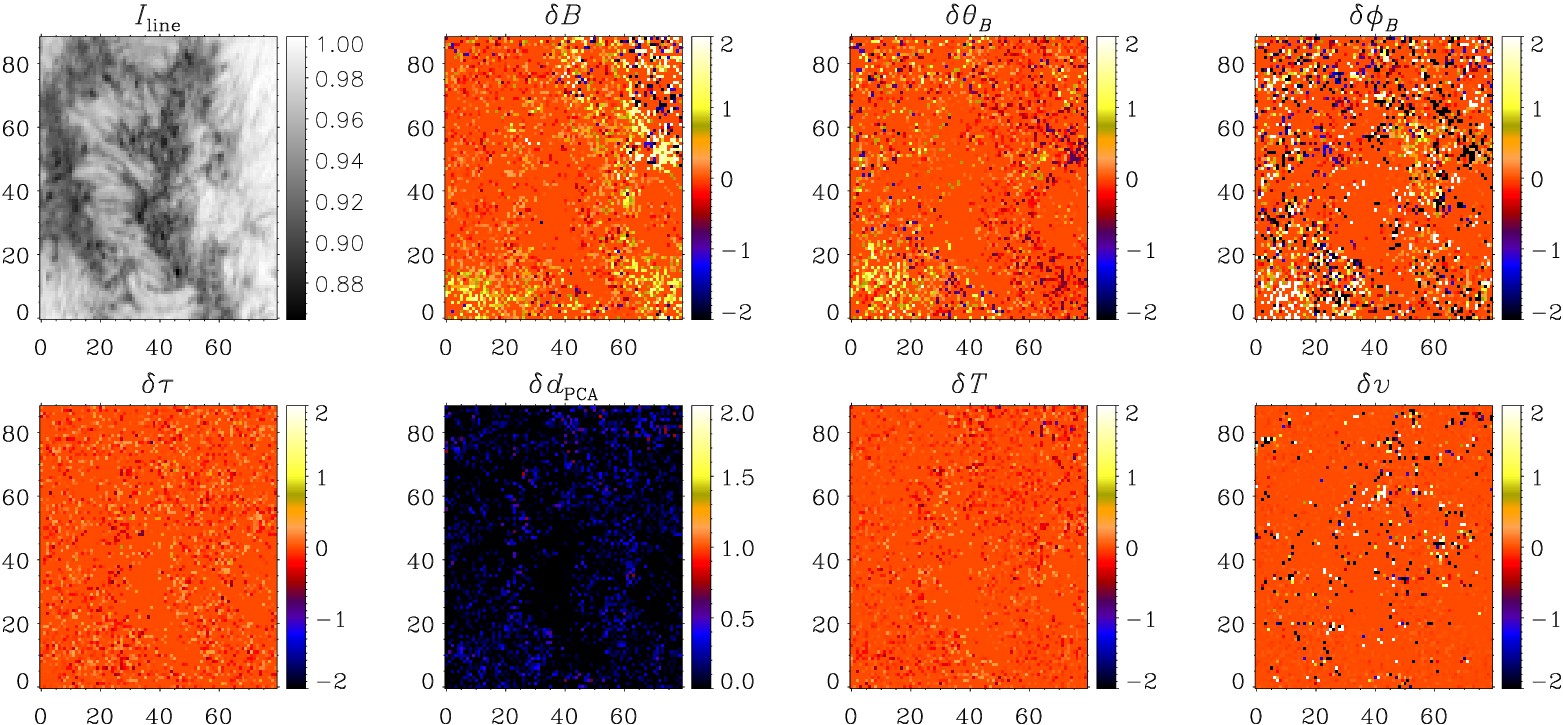}\vspace{8pt}
\includegraphics[width=.8\hsize]{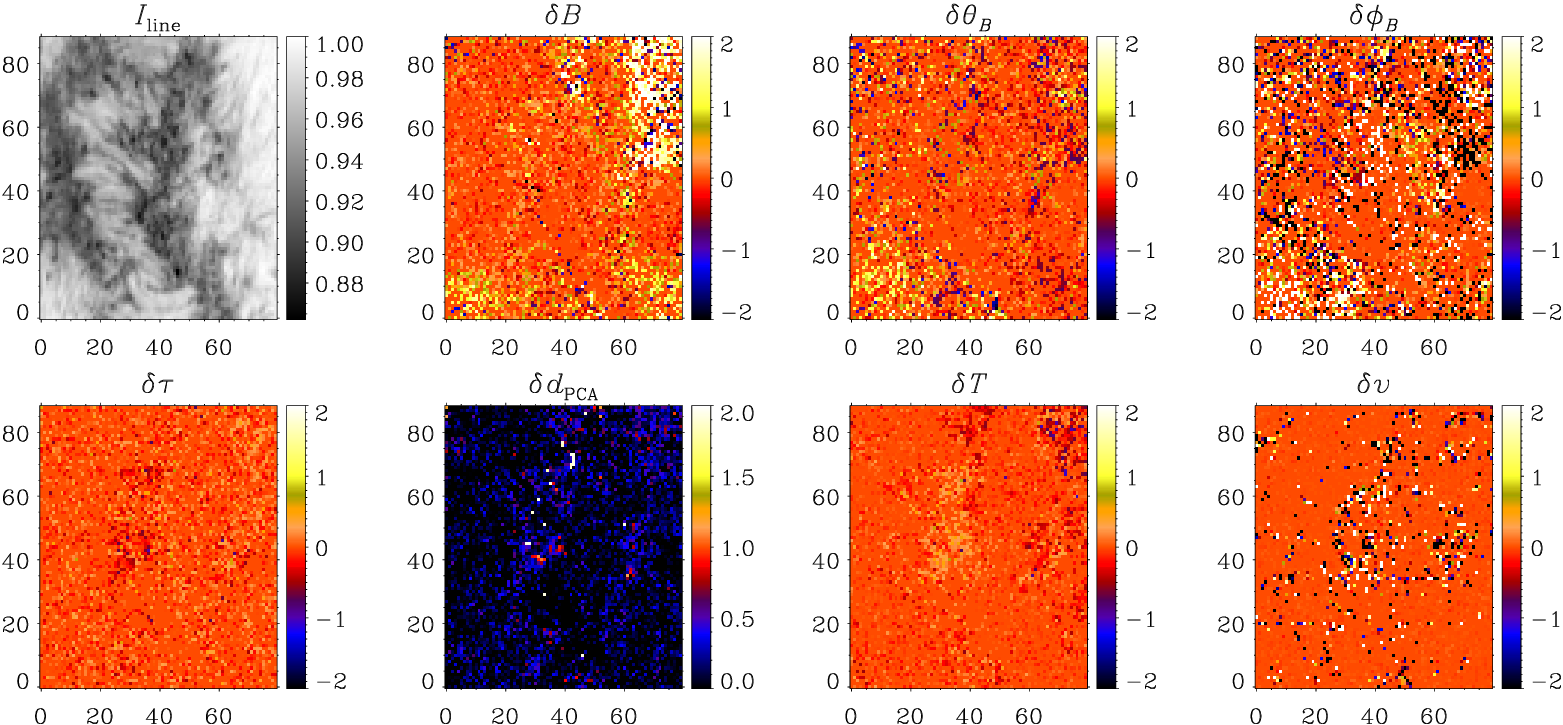}
\caption{\label{fig:difference}
Difference maps for the indexed inversion of 1st (top) and 2nd order
(bottom), relative to the inversion with the non-indexed database.
For each inverted parameter, $p$, these maps show the relative
%difference $\delta p\equiv(p^{(k)}-p^{(0)})/p^{(0)}$, where $k$ is 
difference $\delta p\equiv 2(p^{(k)}-p^{(0)})/(|p^{(k)}|+|p^{(0)}|)$, 
where $k$ is 
the indexing order of the inversion database.}
\end{figure}

\section{Test results and discussion} \label{sec:results}

We tested the proposed method of indexing of the PCA inversion
database to a set of \ion{He}{1} 1083\,nm observations performed by
one of the authors (B.L.) with the Tenerife Infrared Polarimeter~II 
instrument (TIP~II; \citealt{Co07}) deployed at the German Vacuum 
Tower Telescope (VTT, Tenerife, Spain). 
% (Bruce's addition)
These observations were performed on NOAA Active Region 11259 (406\,arcsec 
east, 334\,arcsec north of disk center) on 22 July, 2011 between 
07:31 and 08:24 UT, with full Stokes spectral imaging of the region 
around the \ion{He}{1} 1083\,nm lines from TIP~II, plus imaging of the 
spectral region around the \ion{Ca}{2} lines at 854.2\,nm in Stokes~$I$ 
only with the VTT spectrograph. Atmospheric seeing was very good 
during these 
observations, and the image quality was enhanced by usage of the 
KAOS adaptive optics system. Here we discuss only the observations of 
the \ion{He}{1} lines. The spectral sampling of the TIP~II data was 
0.0109\,\AA/pixel, spanning a spectral region of 11\,\AA\ around 
the \ion{He}{1} lines including the \ion{Si}{1} line at 1082.7\,nm.  
Along the slit dimension (solar N-S) the sampling was 0.175\,arcsec, 
spanning 78\,arcsec.  At each of the 240 scan positions (from solar E to W) 
of the spectrograph slit the signal was integrated for 10 seconds. 
During data reduction the data were re-binned by a factor of 3 
in wavelength, a factor of 5 along the slit, and a factor of 3 in the 
slit scan direction in order to increase the signal-to-noise ratio 
(S/N) for the weak Stokes polarization signals. The data subjected to 
this analysis has 336 wavelength steps of 0.0328\,\AA, 89 positions 
along the slit of spacing 0.875\,arcsec, and 80 positions in the slit 
scan direction of 1.05\,arcsec. The resulting S/N as determined 
empirically from the r.m.s.\ fluctuation in the polarization 
continua is 0.024\%, 0.018\%, and 0.028\%, respectively for $Q$, $U$,
and $V$.
%\textbf{%
The spectral range of these observations extends well beyond the blue
and red wings of \ion{He}{1} 1083\,nm. As demonstrated by the Stokes
eigenprofiles of Figure~\ref{fig:eigenprofiles}, the 151 wavelength 
points adopted for the PCA database are sufficient to encompass the
spectral range of the multiplet.%}

We inverted all 7120 pixels in the map, first with the
original database of 0.75 million models, and then with the same 
database indexed according to the proposed strategy, using $n=1$ 
(16 partition classes) and $n=2$ (256 partition classes). 
	Following the argument given in the previous
section, with regard to the maximum number $k_0$ of orders to retain for
the inversion, the set of variance plots produced for the observed data
indicated that we should use 4 eigenprofiles for Stokes~$I$, and
3 for all of Stokes~$Q$, $U$, and $V$.
	The magnetic maps resulting
from these inversions are displayed in Figure~\ref{fig:results}. 
%For these inversions, we used only 0.5 million models, out of the total
%0.75 million. 
The inversion times for the three tests are given in
Table~\ref{tab:results}. 

The inversion results shown in Figure~\ref{fig:results} are presented in
the form of ``magnetic maps'', each consisting of eight panels. These, 
from left to right and top to bottom, give the line-center 
intensity map of the observed region, the magnetic field strength in 
gauss, the magnetic field vector inclination from the local vertical, 
and its azimuth counted counterclockwise from the direction defined by 
the projected solar radius through the observed point (not showing in
these maps), the line-center optical depth of the slab from which the 
line's Stokes profiles emerge, the longitudinal component of the magnetic 
field, the plasma temperature as defined by the line's Doppler width, 
and finally the line-of-sight velocity as determined by the line's 
Doppler shift, where the zero reference is given by the line position 
averaged over the entire map.

\begin{figure}[!t]
\centering
\includegraphics{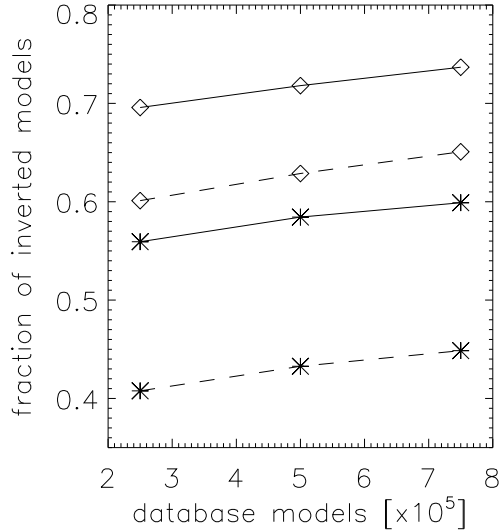}
\caption{\label{fig:stat}
Plots of the percentage of inverted models whose properties are
preserved in passing from a non-indexed to an indexed inversion: ($*$)
all physical parameters of the models are preserved; ($\diamond$) 
the inverted magnetic 
strength varies by less than 20\%. The continuous (dashed) lines 
correspond to the indexed inversions of order 1 (2).} 
\end{figure}

The differences introduced in the inversion by the indexing of the
PCA database are shown in Figure~\ref{fig:difference}, for both orders of
indexing. In those maps, the $B_{\rm LOS}$ panel has been replaced
by the relative increase of the PCA distance, which estimates the
goodness of the inversion. Because of the introduction of disjoint
classes in the database, we must expect that the PCA distance can only
increase (hence, leading to an overall worse fitting of the
observations), with respect to the case where the PCA database is not
indexed. For each of the inverted parameters, $p$, the maps of 
Figure~\ref{fig:difference} give the relative difference 
%$\delta p\equiv(p^{(k)}-p^{(0)})/p^{(0)}$, where $k$ is the indexing 
$\delta p\equiv 2(p^{(k)}-p^{(0)})/(|p^{(k)}|+|p^{(0)}|)$, where 
$k$ is the indexing 
order of the PCA database used for the inversion.
These maps show that the results for the indexed inversion of order~1 
(2) are exactly identical to those for the non-indexed inversion only 
for 59.9\% (44.9\%) of the inverted points. As an example of the changes 
produced by the indexing of the 
PCA database, we consider the case of the magnetic field strength. 
This is found to change by less than 20\% over 73.7\% (65.1\%) 
of the observed region, for the indexed inversion of order 1 (2).
(See also Fig.~\ref{fig:stat}).

We notice, however, that the largest relative errors on the inverted field 
strength, with variations in excess of 100\%, tend to occur in regions 
of the map where the inferred magnetic field is very small 
(cf.~maps in Figure~\ref{fig:results}).
This is a direct consequence of our definition of the relative error 
$\delta p$ given above. While this definition was adopted so to prevent
the inversion error from diverging in some points of the map, it is 
evident that large errors can still be expected for small values of 
$p^{(0)}$, with a theoretical maximum of 200\% (cf.~maps in
Figure~\ref{fig:difference}) when $p^{(0)}$ vanishes.
We also observe that several points along the magnetic neutral
line suffer a noticeable increase of the error in the inferred magnetic
strength for the order 2 of database indexing. While the number of such 
points is by no means statistically significant, it is easy to provide
an interpretation of this result. Along the neutral line, we expect that 
the PCA coefficients associated with Stokes $V$ will be very close to 
zero, and therefore the signs of those coefficients, which determine the
specific indexing class of those profiles, loses significance. Reliance 
on the sign of those coefficients is therefore bound to increase the 
inversion error, and this will be more noticeable when the number of 
models in each class is smaller, as it is the case for the order 2 of 
database indexing.

It is expected that improving the statistical significance of the 
inversions, by increasing the overall number of models in the 
database, will reduce the difference between non-indexed and 
indexed inversions. 
This well illustrated by Figure~\ref{fig:stat}, which shows
how the percentage of inverted models that preserve a given set of 
properties, in passing from non-indexed to indexed inversions, changes 
as a function of the size of the
inversion database. For this figure, we have considered the two cases
mentioned earlier: one where \emph{all} values of the physical parameters
of the model are exactly preserved by the indexing of the inversion
database (star symbols), and the other where we look at the percentage 
of models where the inferred value of the magnetic strength is found to 
change by less than 20\% (diamond symbols). The continuous (dashed) curve 
shows the case for the indexed inversion of order 1 (2). This figure
summarizes the obvious fact that increasing the indexing order also
increases the inversion error, because of the reduced number of database
models falling in each class. At the same time, it also shows that, by
increasing the number of models in the database, the inversion errors
are also bound to decrease. For example, the trend shown in 
Figure~\ref{fig:stat} suggests that the errors on the inferred magnetic
strength attained using a database with 1.2-1.3M models with an indexed 
inversion of order 2 should be comparable to the errors from an indexed 
inversion of order 1 over a database with only 0.25M models.
%This is demonstrated by the fact that, using 
%only 2/3 of the database models for the inversion, the percentage 
%of results that are unchanged by the indexing of the PCA database 
%decreases to 58.4\% (43.3\%), for the case of the indexed inversion 
%of order 1 (2). The percentage of the map where the magnetic field 
%strength is found to change by more of 20\% increases instead
%to 28.2\% (37.1\%). 
%This trend is confirmed by the inversion results that we obtain 
%if we use only 1/3 of the database models, in which case 
%the percentage of the map where the inversion results 
%are unchanged by the indexing of the PCA database goes down to 55.9\% 
%(40.8\%), for the the case of the indexed inversion of order 1 (2),
%whereas the magnetic field strength changes by more than 20\% 
%over 30.4\% (39.9\%) of the map. 
%
These results indicate that the stability of the inversion results
under indexing of the PCA database depends in fact on the density of
the models in the database, although it changes rather slowly as a
function of the total number of models.

\begin{figure}[t!]
\centering
\includegraphics[width=.49\hsize]{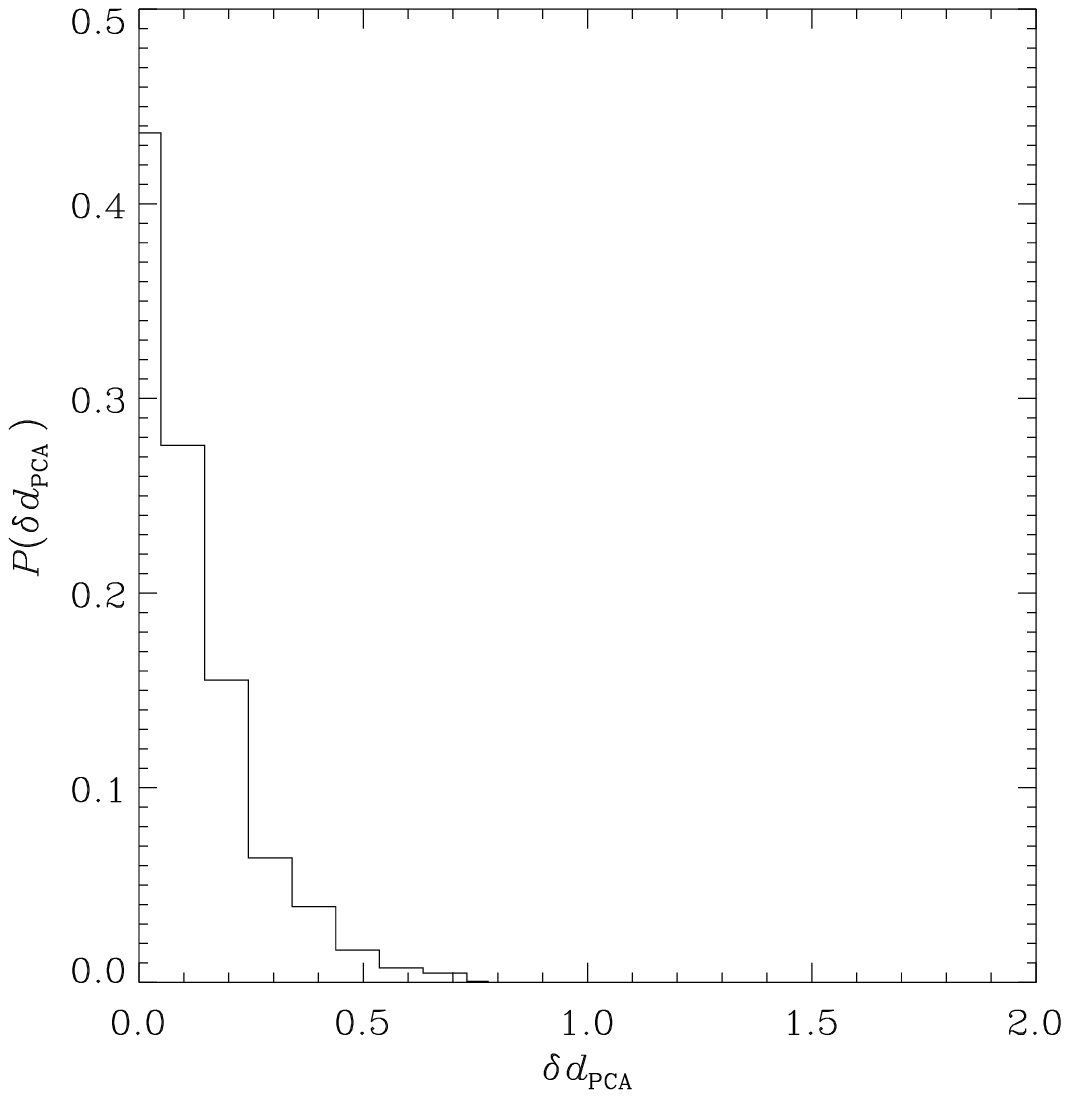}\kern 8pt
\includegraphics[width=.49\hsize]{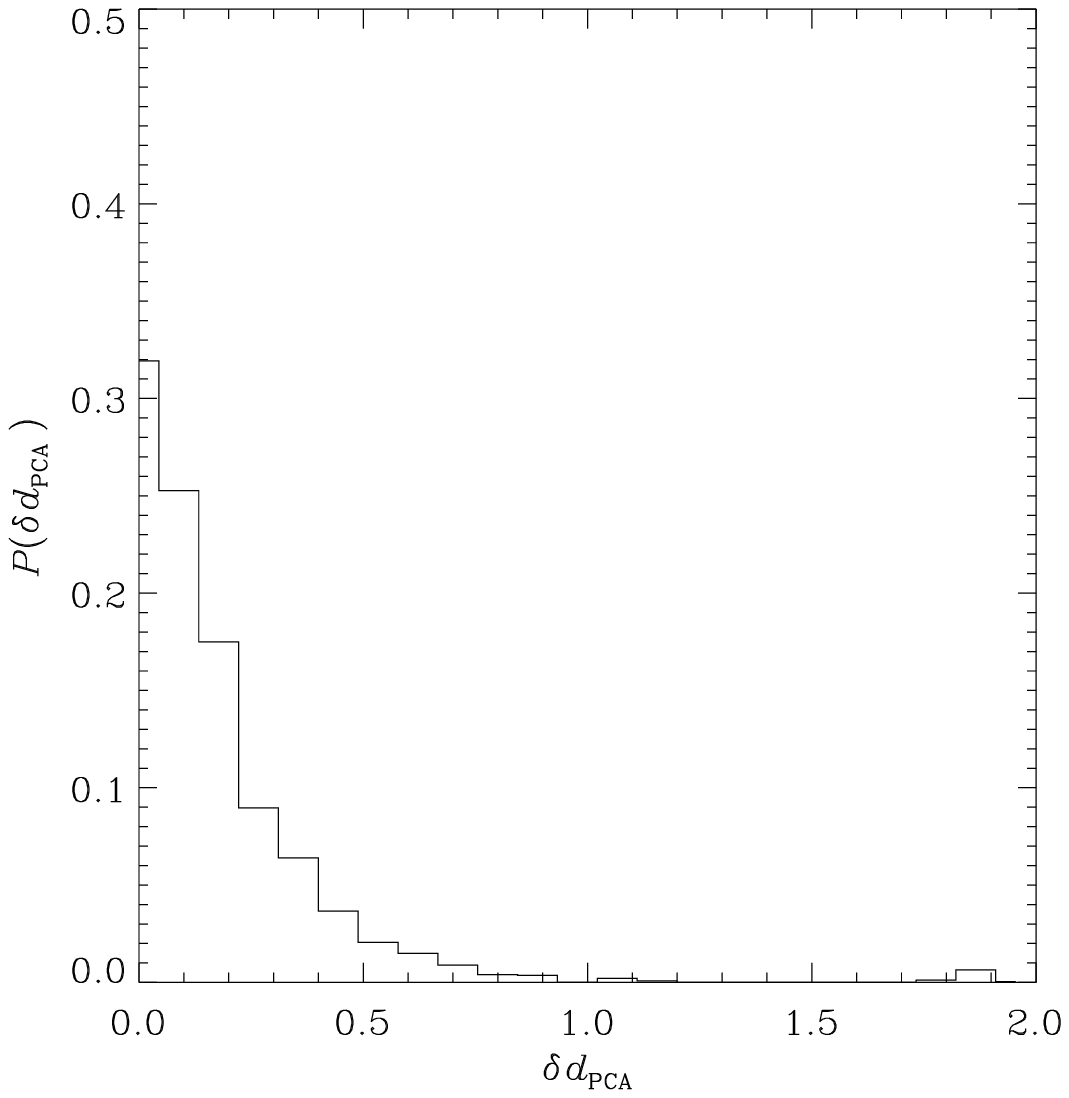}
\caption{\label{fig:PCAdistr}
Normalized distributions of the relative change of the PCA distance, 
$\delta d_{\rm PCA}$, for the indexed inversions of order 1 (left) 
and 2 (right). These
distributions are calculated on the subset of map points for which the
magnetic field strength is found to vary by more than 20\% in passing
from non-indexed to indexed inversions.}
%For these plots, we compared
%inversions that were based on the full PCA database of 0.75 million
%models.}
\end{figure}

It is legitimate to question whether the changes in the inferred values 
of the model parameters, which occur in passing from non-indexed to 
indexed inversions, may affect too large a portion of the map to 
justify the proposed method as a reliable approach to 
spectro-polarimetric inversion. On the other hand, the observed 
changes must be interpreted in the light of the possible presence 
of intrinsic ambiguities in the line formation model, which can 
result in very similar sets of emerging Stokes profiles even for 
magnetic configurations that may differ significantly. This fact is 
well illustrated by the distribution of $\delta d_{\rm PCA}$ for 
the set of map points where the inferred value of the magnetic field 
strength changes by more than 20\% between non-indexed and indexed
inversions (see also Figure~\ref{fig:difference}). 
Figure~\ref{fig:PCAdistr} shows this (normalized) 
distribution for inversions based on the full databases of 0.75 
million models. The left (right) panel shows the increase of the PCA
distance for the indexed inversion of order 1 (2). The fact that this
distribution gathers decidedly around zero, with 72\% (62\%) of the
models showing less than a 20\% increase in the PCA distance,
statistically demonstrates that the changes in the inferred magnetic
field caused by the indexing of the inversion databases has only a minor
effect on the goodness of the profile fit, and that those changes are
then compatible with the presence of intrinsic ambiguities of the line 
formation model.

The proposed method for the indexing of PCA inversion databases is
particularly easy to implement. Along with the manyfold increase
in the inversion speed that is possible to attain, this is another 
attractive feature of the method. As it is apparent from comparing 
qualitatively the maps of Figure~\ref{fig:results}, as well as from 
the more detailed statistical analysis of the variations produced 
with different orders of indexing, which we presented above, the 
proposed method appears to be adequate for fast handling of large 
synoptic datasets, such as those from full-disk observations of the 
Sun. Instruments such as the Synoptic Optical Long-term Investigations 
of the Sun (SOLIS) of the National Solar Observatory \citep{Ke98}, or the 
Chromosphere Magnetometer (ChroMag) of the High Altitude 
Observatory \citep{DW12}, currently under testing, can profit greatly 
from the proposed strategy of spectro-polarimetric inversion, and
represent ideal testbeds of the method. Using 
a database with 0.5 million models, and an order of indexing of 2, it 
would take only about 30 minutes to fully invert an observation of the 
entire solar disk with a 1\,arcsec spatial resolution.

On the other hand, the dramatic increase of the inversion speed 
granted by database indexing realistically opens to the possibility
of using much larger PCA databases than in the past. This would allow
to perform high-precision spectro-polarimetric inversions of smaller 
regions of the Sun, the typical size of a medium active region 
($\sim 4\,\rm arcmin^2$), using PCA databases with several tens of 
millions of models, for which the downsides of database indexing
that we have previously discussed are expected to be significantly
reduced. Another possible application of PCA indexed inversion
is to provide a fast initialization of spectro-polarimetric inversions 
that rely on elaborate optimization schemes, such as the
Levenberg-Marquardt algorithm \citep[e.g.,][]{AR08}, as an alternative 
to more cumbersome initialization methods such as those based on the 
genetic algorithm.

%These differences can be expected to vanish eventually, in the limit 
%of very large databases, only if both the systematic and random errors 
%in the observations lie below the sensitivity threshold set by the 
%PCA coefficients associated with the order of indexing of the database.

\acknowledgments The authors thank the Kiepenheuer-Institut
f\"ur Sonnenphysik for a grant of observing time that permitted the
observations reported herein to be obtained.  We also thank C. Beck, M.
Collados, C. Kuckein, R. Rezeai,  and W. Schmidt for assistance during
the observing run, and C. Kuckein for assistance in data reduction.
We thank HAO colleagues G.\ de Toma and A.\ Skumanich for a careful 
reading of the manuscript and for helpful comments. We are deeply
indebted to the anonymous referee, who has done a very scrupulous
job in reviewing the manuscript, and in suggesting clarifications and
improvements to the original presentation of this work.

\end{document}